\documentclass[12pt]{article}
\usepackage{graphicx}
\usepackage{subfigure}

\newcommand\pubnumber{SNSN-323-63}
\newcommand\pubdate{\today}\newcommand{\comment}[1]{\par\noindent {\em\small [#1]}}
\renewcommand{\comment}[1]{}

\newcommand{\unit}[1]{\ensuremath{\rm\,#1}}

\newcommand{\invfb}{\unit{fb^{-1}}}

\newcommand{\DGs}{\ensuremath{\Delta\Gamma_{s}}}

\newcommand{\phis}{\ensuremath{\phi_{s}}}

\newcommand{\BAR}[1]{\overline{#1}}

\newcommand{\particle}[1]{{\ensuremath{#1}}}

\newcommand{\Bd}{\particle{B^0}}
\newcommand{\Bs}{\particle{B^0_s}}

\newcommand{\Bsbar}{\particle{\BAR{B}{^0_s}}}

\newcommand{\Jpsi}{\particle{J\!/\!\psi}}

\newcommand{\KS}{\particle{K^0_S}}  
     
\newcommand{\BsKKp}{\decay{\Bs}{\Jpsi\KS K \pi}}

\newcommand{\BdKpp}{\decay{\Bd}{\Jpsi\KS\pi\pi}}

\newcommand{\BdKKK}{\decay{\Bd}{\Jpsi\KS K K }}

\newcommand{\BdKshh}{\decay{\Bd_{(s)}}{\Jpsi\KS h^{(')+} h^- }}     

\newcommand{\Kst}{\particle{K^{*0}}}  
  
\newcommand{\Kstbar}{\particle{\BAR{K}^{*0}}}  
\newcommand{\BsJKbar}{\decay{\Bs}{\Jpsi \Kstbar}}
\newcommand{\BdJK}{\decay{\Bd}{\Jpsi \Kst}}

\newcommand{\decay}[2]{\particle{#1\!\to #2}}

\newcommand{\BsJphi}{\decay{\Bs}{\Jpsi\phi}}             
\newcommand{\BsJpipi}{\decay{\Bs}{\Jpsi\pi^-\pi^+}}

\newcommand{\BsJKs}{\decay{\Bs}{\Jpsi \particle{K^0_S}}}
\newcommand{\BdJKs}{\decay{\Bd}{\Jpsi \particle{K^0_S} }}

\newcommand{\BsJKK}{\decay{\Bs}{\Jpsi K^- K^+}}
           
\newcommand{\beq}{\begin{equation}}
\newcommand{\eeq}{\end{equation}}

\def\napoli{CPPM, Aix-Marseille Universit\'e CNRS/IN2P3, Marseille, France}

\def\Title#1{\begin{center} {\Large #1 } \end{center}}
\def\Author#1{\begin{center}{ \sc #1} \end{center}}
\def\Address#1{\begin{center}{ \it #1} \end{center}}

\newcommand\pubblock{\rightline{\begin{tabular}{l} \pubnumber\\
         \pubdate  \end{tabular}}}
\newenvironment{Abstract}{\begin{quotation}  }{\end{quotation}}
\newenvironment{Presented}{\begin{quotation} \begin{center} 
             PRESENTED AT\end{center}\bigskip 
      \begin{center}\begin{large}}{\end{large}\end{center} \end{quotation}}
\def\Acknowledgements{\bigskip  \bigskip \begin{center} \begin{large}
             \bf ACKNOWLEDGEMENTS \end{large}\end{center}}

\begin{document}
\begin{titlepage}
\pubblock

\vfill
\Title{Measurement of CP violating phase $\phi_s$ and control of penguin pollution at LHCb}
\vfill
\Author{Walaa KANSO\\{\bf on behalf of the LHCb collaboration} }
\Address{\napoli}
\vfill
\begin{Abstract}
The study of CP violation in \Bs\, oscillations is a key measurement at the LHCb experiment. In this document, we discuss the latest LHCb results on the CP-violating phase, called $\phi_s$, using \BsJKK\, and \BsJpipi\, channels. To conclude on the presence of New Physics in $\phi_s$, the estimation of the sub-dominant contributions from the Standard Model becomes crucial now. We outline a method to estimate the contribution of penguin diagrams in $\phi_s$. Branching fractions and upper limits of \BdKshh\,($ h^{(')}=K,\pi)$\, modes are presented.
\end{Abstract}
\vfill
\begin{Presented}
8th International Workshop on the CKM Unitarity Triangle\\
(CKM 2014)\\
8-12 September 2014, Vienna, Austria\\
\end{Presented}
\vfill
\end{titlepage}
\def\thefootnote{\fnsymbol{footnote}}
\setcounter{footnote}{0}
%

\section{Introduction}

The interference between \Bs\, mesons decaying directly via $b\to c\overline{c}s$ transitions to CP eigenstates and those decaying after \Bs-\Bsbar\, oscillations gives rise to a CP violating phase called $\phi_s$.


 
Within the Standard Model, the decay can occur via two main topologies: the predominant tree topology and the sub-leading penguin diagram. 
New Physics processes, e.g., new particles contributing to the box diagrams, can modify the value of $\phi_s$:
\begin{center}
$\phi^{\rm meas}_{s} = -2\beta_s+\Delta\phi_{s}^{\rm peng}+\delta^{\rm NP}$
\end{center}
The indirect determination via global fit to experimental data gives:\\
$-2\beta_s=2\arg(\frac{V_{ts} V_{tb}^{\star}}{V_{cs} V_{cb}^{\star}}) =-0.0363\pm0.0013$~\cite{J.Charles}. The theoretical uncertainty on $\phi_s$ is mainly due to unknown penguin contributions $\Delta\phi_{s}^{\rm peng}$. The control of penguin pollution is limited by large theoretical uncertainties. Thus, experimental measurements are very useful to constrain the contribution of penguin diagrams. 
Therefore, we should estimate $\Delta\phi_{s}^{\rm peng}$, otherwise we may incorrectly interpret the Standard Model penguin contributions as signs of New Physics. 
In Section 2, we summarize the measurement of the CP-violating phase $\phi_s$ in \BsJKK. In Section 3, we report on a recent update of the same measurement using \BsJpipi. 
The studies of penguin pollution in $\phi_s$ and 2$\beta$ are described in Section 4 and 5 respectively. A recent result on the branching ratio of \BdKshh\,($ h^{(')}=K,\pi)$\, modes is shown in Section 6.

\section{\BsJKK\,}

The purpose of this analysis is to mainly measure $\phi_s$, the decay width difference $\Delta\Gamma_s$ and $\Gamma_s$. 
$\Bs \to \Jpsi[\to \mu^+ \mu^-]\phi[\to K^+ K^-] $ is pseudo-scalar decaying into two vector mesons.
The study of this decay requires an angular analysis in order to disentangle CP-odd/CP-even mixture of the final state. After the statistical background subtraction~\cite{xie}, a fit to decay time ($t$) and three angles in helicity frame ($\Omega= \theta_{\mu}, \theta_K, \varphi_h$) is performed in six bins of $m_{KK}$~\cite{P.R.D}.
To account for the detector and selection effects, the decay time acceptance is taken from real data and the angular acceptance is studied in the simulated data. The finite decay time resolution is modelled by a single Gaussian of width $S_{\sigma_t}\times\sigma_t$, where $\sigma_t$ is the estimated per event decay time uncertainty, and the scale factor, $S_{\sigma_t}$, is measured in a sample of prompt $J/\psi\rightarrow \mu^+\mu^-$. The effective resolution is 45 fs for \BsJphi. The $B^{0}_s$ flavour, at production, is determined by the combination of the same side kaon tagger and the opposite side taggers, which gives the overall tagging power: $\epsilon(1-2\omega)^2=(3.13 \pm 0.23)\% \nonumber$, where $\epsilon$ indicates the tagging efficiency, and $\omega$ the mistag rate~\cite{Aaij:2013oba}. Using 1 \invfb\, of real data collected in 2011, LHCb obtained:
\begin{center}
$
  \begin{array}{ccllllllll}
    \phi_s &\;=\; & 0.07  &\pm & 0.09  & \rm{\scriptsize(stat)} &\pm & 0.01 & \rm{\scriptsize(syst)}\\
    \Gamma_s &\;=\; & 0.663  &\pm & 0.005 & \rm{(stat)} &\pm & 0.006 & \rm{(syst)}\\
    \DGs    &\;=\; & 0.100   &\pm & 0.016    & \rm{(stat)} &\pm & 0.003 & \rm{(syst)} \\
  \end{array}
$
\end{center}











\section{\bf{\BsJpipi}}
\begin{figure}[htbp]
\centering
 \subfigure[]{
 \includegraphics[width=0.4\linewidth]{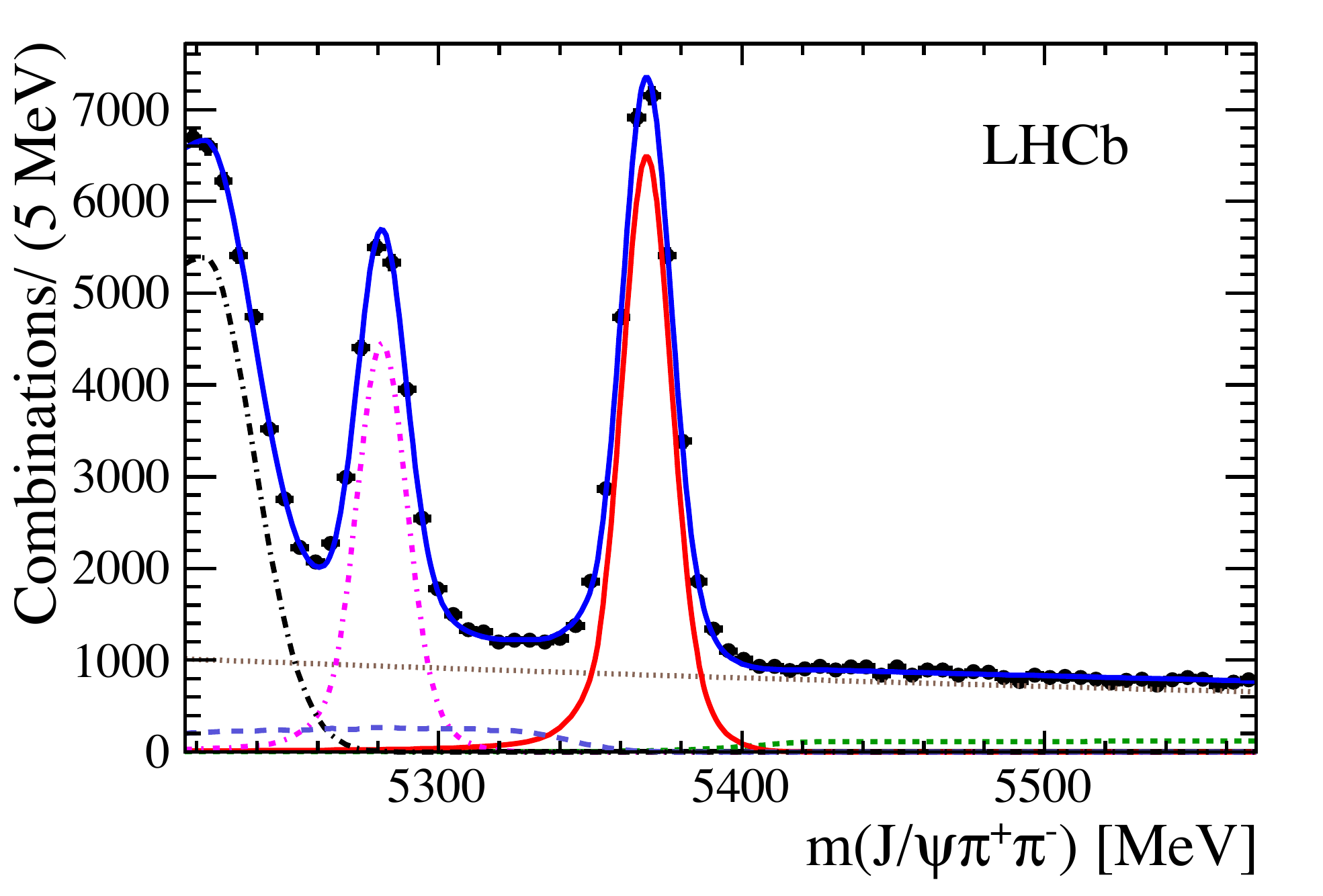}
 \label{fig:pipiMass}
 }
 \subfigure[]{
 \includegraphics[width=0.4\linewidth]{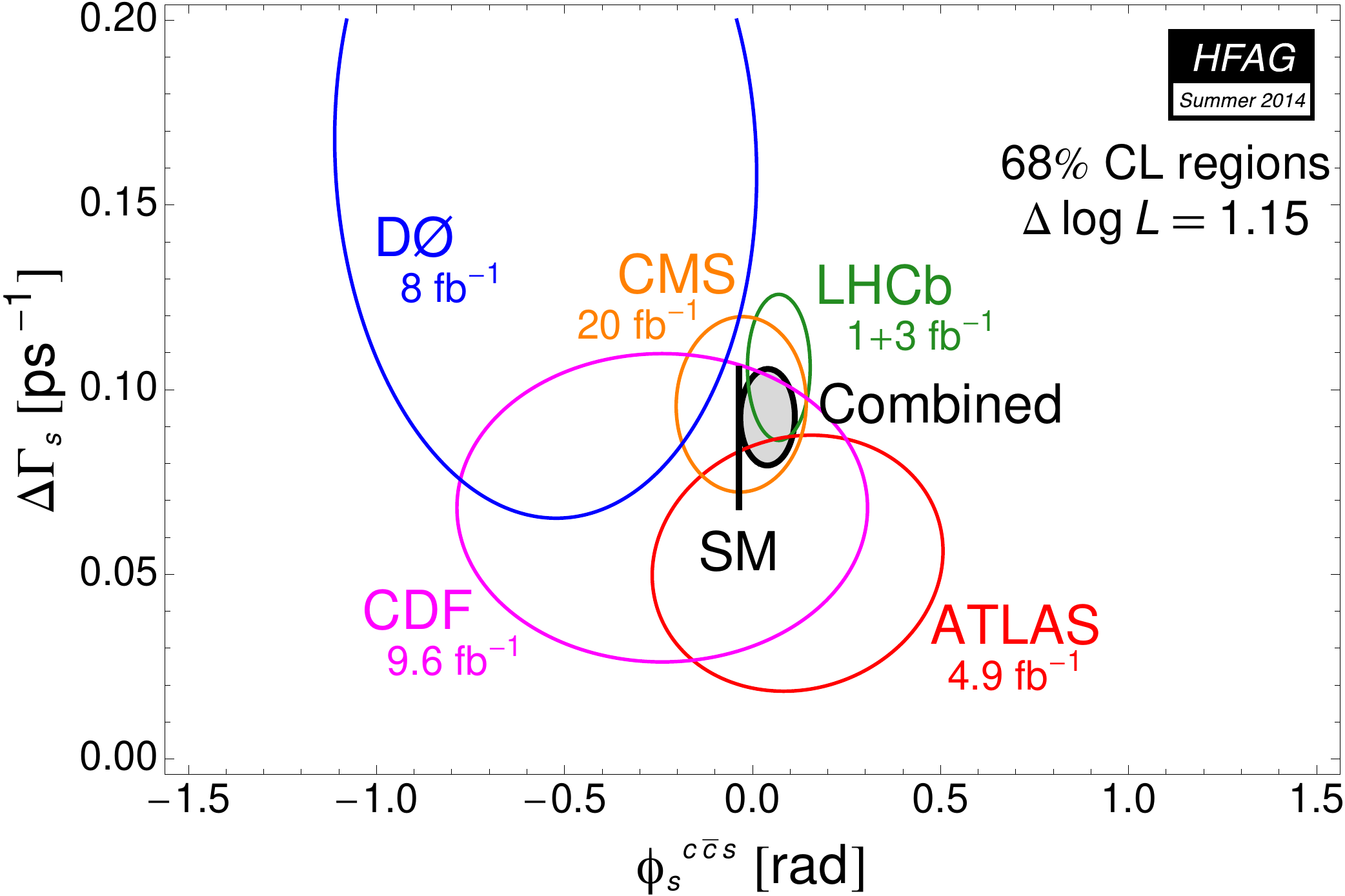}
 \label{fig:DGsVersusphisSummer2014}
 }
\caption{(a): Invariant mass of $J/\psi \pi^{\pm} \pi^{\mp}$. The (red) solid line shows the \Bs\,, the the (brown) dotted line shows the exponential combinatorial background, the (green)
short-dashed line shows the $B^{\pm}$ background, the (magenta) dot-dashed line shows the \Bd\, signal, the (light blue) dashed line represents some misreconstructed decays, the (black) dot-dashed
line is the \Bd$ \rightarrow J/\psi K^{\pm} \pi^{\mp}$ reflection and the (blue) solid line is the total. (b): Unofficial overview of the current experimental constraints in the $\phi_s$-$\Delta\Gamma_s$ plane, modified compared to the original ~\cite{HFAG2012} to include these results as well as the summer 2014 updates from Atlas~\cite{Aad:2014cqa} and CMS~\cite{CMS:2014jxa}.}

\end{figure}

\begin{figure}[htb]
\centering
\includegraphics[width=0.3\linewidth]{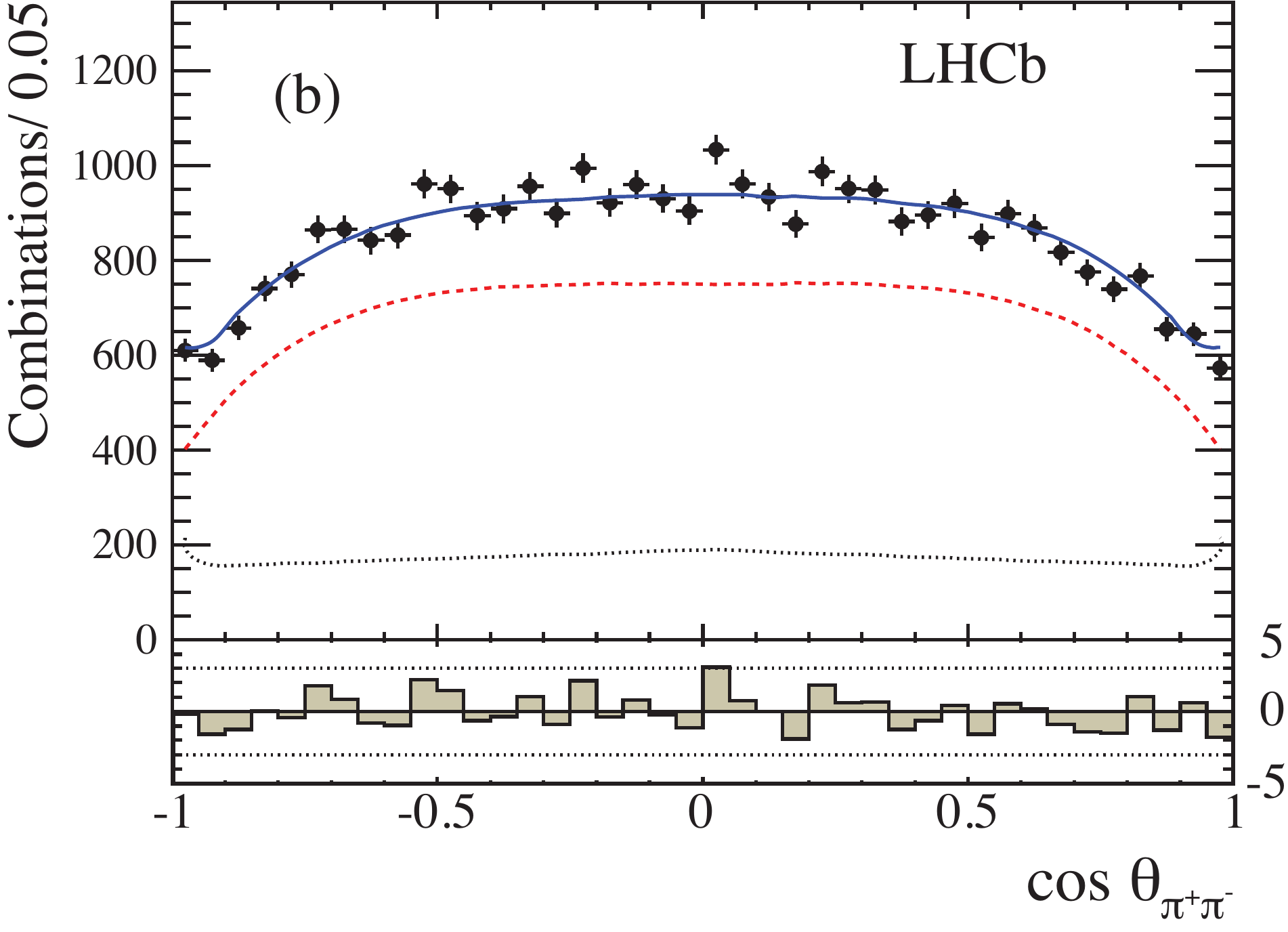}
\includegraphics[width=0.3\linewidth]{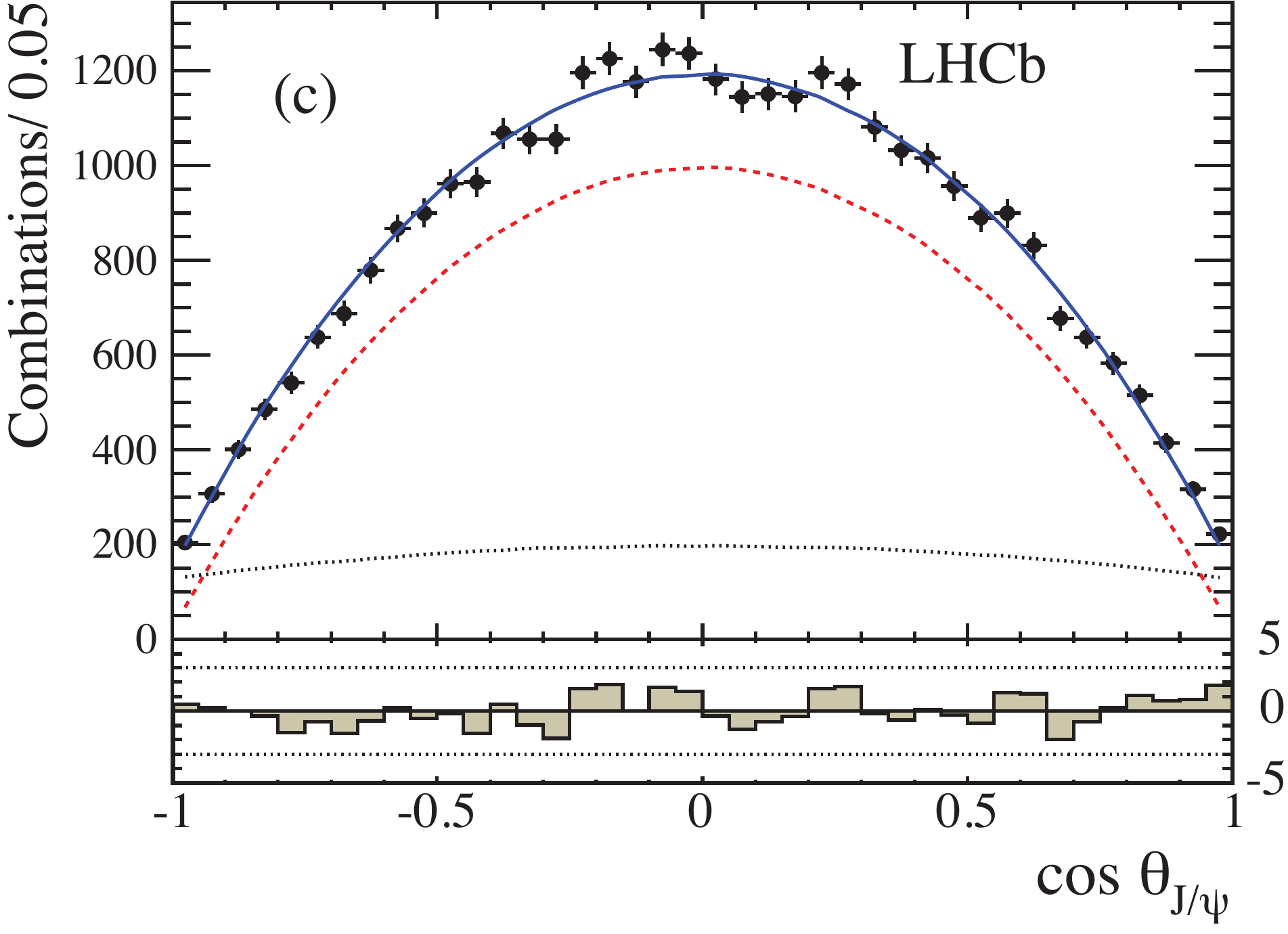}
\includegraphics[width=0.3\linewidth]{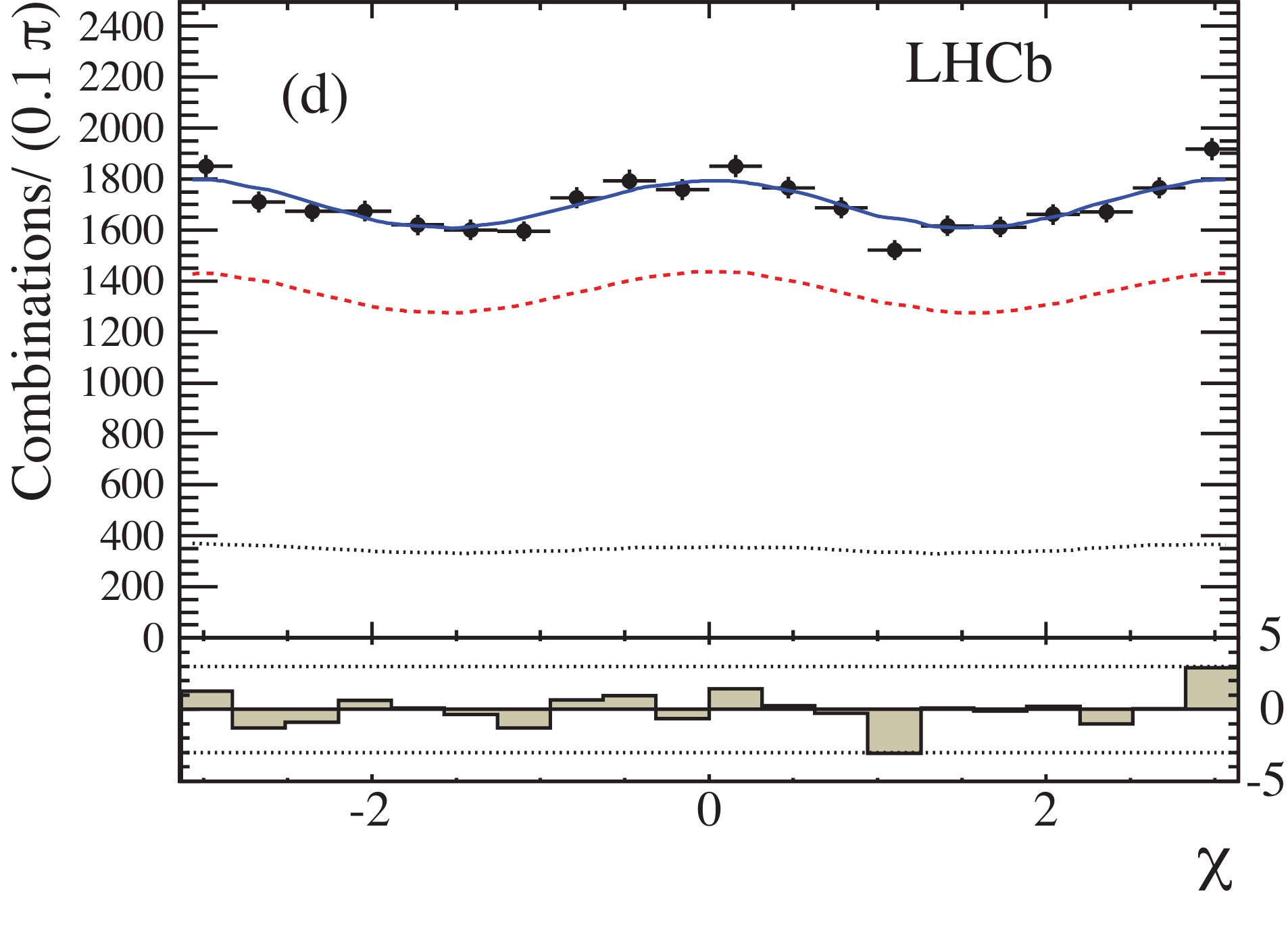}
\caption{Projections of the three decay angles. The signal fits are shown with (red) dashed lines, the background with a
(black) dotted lines, and the (blue) solid lines represent the total fits. The difference between
the data and the fits divided by the uncertainty on the data is shown below.}
\label{fig:pipi}
\end{figure}

Another useful channel to measure $\phi_s$ is \BsJpipi. This decay is 97.7$\%$ dominated by the CP$-$odd component, at 95 $\%$  C.L~\cite{PhysicsLetters}.  
Nevertheless, an angular analysis is needed to obtain the small CP-even component. We performed a six-dimensional fit of \Bs\, and $\pi\pi$ masses, time and the three helicity angles ($m_\Bs$ , $t$, $m_{\pi\pi}$, $\Omega$).
The decay time acceptance is determined in $\BdJK$ channel and simulation, while $\Delta\Gamma_s$ and $\Gamma_s$ are taken from \BsJphi. 
This analysis aims to measure $\phi_s$ and $|\lambda|$, related to the direct CP violation. 
By following the same methodology as in \BsJphi\, channel to determine the effective decay time resolution, we obtained: 40fs. 
The tagging algorithm achieved a power of:
$\epsilon(1-2\omega)^2=(3.89 \pm 0.25)\% \nonumber$.
It is found that five interfering $\pi^+\pi^-$ states are required to describe the data.
These include the dominant $f_0(980)$, as well as the $f_0(1500)$, $f_0(1790)$, $f_2(1270)$, $f'_2(1525)$.
The resulting decomposition of the $\pi^+\pi^-$ invariant mass spectrum is shown in (fig.~\ref{fig:pipiMass}).
The projections of angular distributions are shown in fig.~\ref{fig:pipi}.


LHCb analysed 3 \invfb\, of \BsJpipi\, collected in 2011 and 2012. Assuming that there is no direct CP-violation ($|\lambda|=1$), the CP violating phase, $\phi_s$, is: 
\begin{center}
$
  \begin{array}{ccllllllll}
 \phi_s &\;=\; & 0.075  &\pm & 0.067  & \rm{(stat)} &\pm & 0.008 & \rm{(syst)}\\
\end{array}
 $
\end{center}

\noindent If the direct CP-violation, represented by $|\lambda|$, is floating, we obtained:
\begin{center}
$
 \begin{array}{ccllllllll}
 \phi_s &\;=\; & 0.070  &\pm & 0.068  & \rm{(stat)} &\pm & 0.008 & \rm{(syst)}\\
 \lambda &\;=\; & 0.89  &\pm & 0.05  & \rm{(stat)} &\pm & 0.01 & \rm{(syst)}\\
\end{array}
 $
\end{center}

\noindent A preliminary combination of 3 \invfb\, \BsJpipi\, with 1\invfb\, \BsJKK, gives: 
\begin{center}
$
 \begin{array}{ccllllllll}
 \phi_s &\;=\; & 0.070  &\pm & 0.054 &\pm & 0.011\\
 
\end{array}
$
\end{center}

\noindent This measurement is consistent with the Standard Model but there is still room for New Physics. A comparison of the LHCb result with those of other experiments is shown in fig.~\ref{fig:DGsVersusphisSummer2014}

\section{Estimating penguin pollution in \phis\, with \BsJKbar\,}
A sample of $\BsJKbar$ is taken as a control channel because the penguin diagrams are not suppressed in this decay. While in the $\BsJphi$ channel, the penguin process is suppressed, by $\lambda^2\sim0.05$, relative to tree diagram. $a_i$ and $\theta_i$ are defined as penguin's parameter for $\BsJphi$~\cite{Faller}. There are two penguin parameters for each of the three polarizations of final states: $i=0,\perp,\|$:
$a_i e^{i\theta_i} = (1-\frac{\lambda^2}{2})\left|V_{ub}/(\lambda V_{cb})\right|\left[\frac{P^i_u+P^i_t}{T^i_c+P^i_c-P^i_t}\right]$.
With $\lambda \sim 0.22$, penguin amplitude is: $P_q , q=u,t,c$, and the tree amplitude is represented by: $T_c$. $a'_i$ and $\theta'_i$ are the penguin parameters for $\BsJKbar$. In order to connect both channels, penguin parameters in $\BsJphi$ and $\BsJKbar$, are supposed equal, using the approximations of SU(3) flavour (quarks $u,d,s$~are identical): 
\begin{center}
$a_i = a'_i,\quad \theta_i = \theta'_i$
\end{center}

\begin{figure}[htb]
\centering
\includegraphics[width=0.5\linewidth]{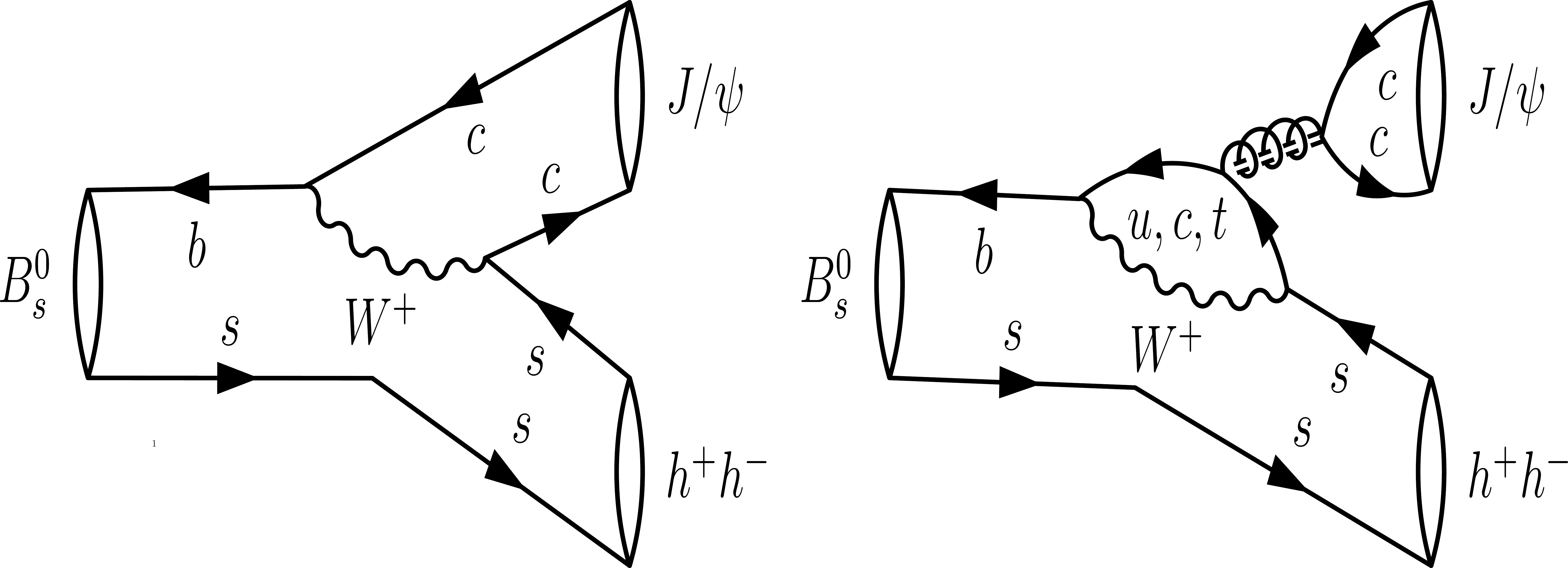}
\caption{Feynman diagrams contributing to the decay $\Bs \to \Jpsi h^⁺h^-$
within the Standard Model. Left: tree; right: penguins.}
\label{fig:treePenguin}
\end{figure}


Therefore, a shift on $\phi_s$ due to penguin diagrams, is calculated for each polarization: 
$\tan(\Delta\phi_{s}^{i,\rm peng}) = \frac{2\epsilon a_i \cos\theta_i \sin\gamma+\epsilon^2 a_i^2 \sin2\gamma}{1+2\epsilon a_i \cos\theta_i \cos\gamma+ \epsilon^2 a_i^2 \cos2\gamma}$, where $\epsilon =\frac{\lambda^2}{1-\lambda^2}$.
From the experimental point of view, there are two observables for each polarization ({\small $i=0,\perp,\|$}): 
The first one, $H_i$,  is proportional to the ratio of branching fractions of the 2 channels, multiplied, respectively, by the corresponding 
polarization fractions. The products of the decay can be in three polarization states: longitudinal,
parallel and perpendicular. This requires an angular analysis to disentangle those states. 
The factor $\left|\frac{\mathcal{A}_i}{\mathcal{A'}_i}\right|^2$ is a critical SU(3) breaking factor calculated theoretically with a big error. 
\begin{equation}
{\scriptsize H_i}=\frac{1-2{a_i} \cos{\theta_i} \cos\gamma+{a_i}^{2}}{1+2\epsilon {a_i} \cos{\theta_i}\cos\gamma +\epsilon^2 {a_i}^2} 
=\frac{1-\lambda^2}{\lambda^2}\left|\frac{\mathcal{A}_i}{\mathcal{A'}_i}\right|^2 \frac{f^i_{J/\psi\Kstbar}.BR(\BsJKbar)}{f^i_{J/\psi\phi}.BR(\BsJphi)}
\label{H}
\end{equation}
The second observable is the direct CP violation in $\BsJKbar$:
\begin{equation}
{{A}^{CP}_{i}}= \frac{2{a_i} \sin{\theta_i}\sin\gamma}{1-2{a_i}\cos{\theta_i}\cos\gamma+{a_i^{2}}}= \frac{\Gamma_{\Bs}^i-\Gamma_{\Bsbar}^i}{\Gamma_{\Bs}^i+\Gamma_{\Bsbar}^i}
\end{equation} 
$H_i$ and ${A}^{CP}_{i}$ form a non trivial system of two equations with two unknowns :
$a_i$ and $\theta_i$ that lead to the shift on $\phi_s$, $\Delta\phi_s^{i,\rm peng}$.
In this channel, an angular analysis is required to calculate the polarization amplitudes. The presence of P and S-wave in the $K\pi$ system has to be studied carefully. 
Using 0.37 \invfb, a fit to mass, angles has been performed and gave~\cite{P.R.D}:\\
$ BR(\BsJKbar)=\left(4.4_{-0.4}^{+0.5} \pm 0.8 \right) \times 10^{-5}$\\
$f_L = 0.50 \pm 0.08 \pm 0.02$, $f_{\parallel} = 0.19^{+0.10}_{-0.08} \pm0.02$\\
In order to determine penguin pollution in \BsJphi, a measurement of the direct CP violation, in \BsJKbar, is needed. To do so, one has to split the data sample into $K^{+} \pi^{-}$ and $K^{-} \pi^{+}$. A direct CP violation measurement and an update, with 3 \invfb\,, of the branching ratio and polarization fractions are ongoing.

\section{Estimating penguin pollution in $2\beta$ with \BsJKs\,}

\BsJKs\, is the ultimate tool for controlling penguin diagrams in \BdJKs\,.
This requires a CP analysis in \BsJKs~\cite{Faller}. LHCb published a previous analysis of \BsJKs\, with 1 \invfb\, of real data~\cite{Nuclear}:\\
$BR(\BsJKs)= \left(1.97\pm 0.23\right)\times10^{-5}$, $\tau^{eff}= 1.75\pm0.12$(stat)$\pm0.07 $(syst) ps\\
A CP analysis is ongoing, as well as an update of the branching ratio with full LHC run 1 data.

\section{\BdKshh\,($ h^{(')}=K,\pi)$\,}

This analysis could help, with high statistics, in the measurement of \phis. It's also interesting for spectroscopy measurements
and the search for exotics~\cite{arXiv:1405.3219}.

New branching fraction measurements of \BdKshh\, ($ h^{(')}=K,\pi$)
have been performed with 1 \invfb\, of real data. 
The channel \BdKpp\, is confirmed and first observations of \BsKKp\, and \BdKKK\,  are published with more than 7$\sigma$. 


In the first place, \BdKpp\, has been measured relative to \BdJKs. Using \BdKpp\, updated branching fraction, all others modes are measured as you can see below:
\begin{center}
\includegraphics[width=0.9\linewidth]{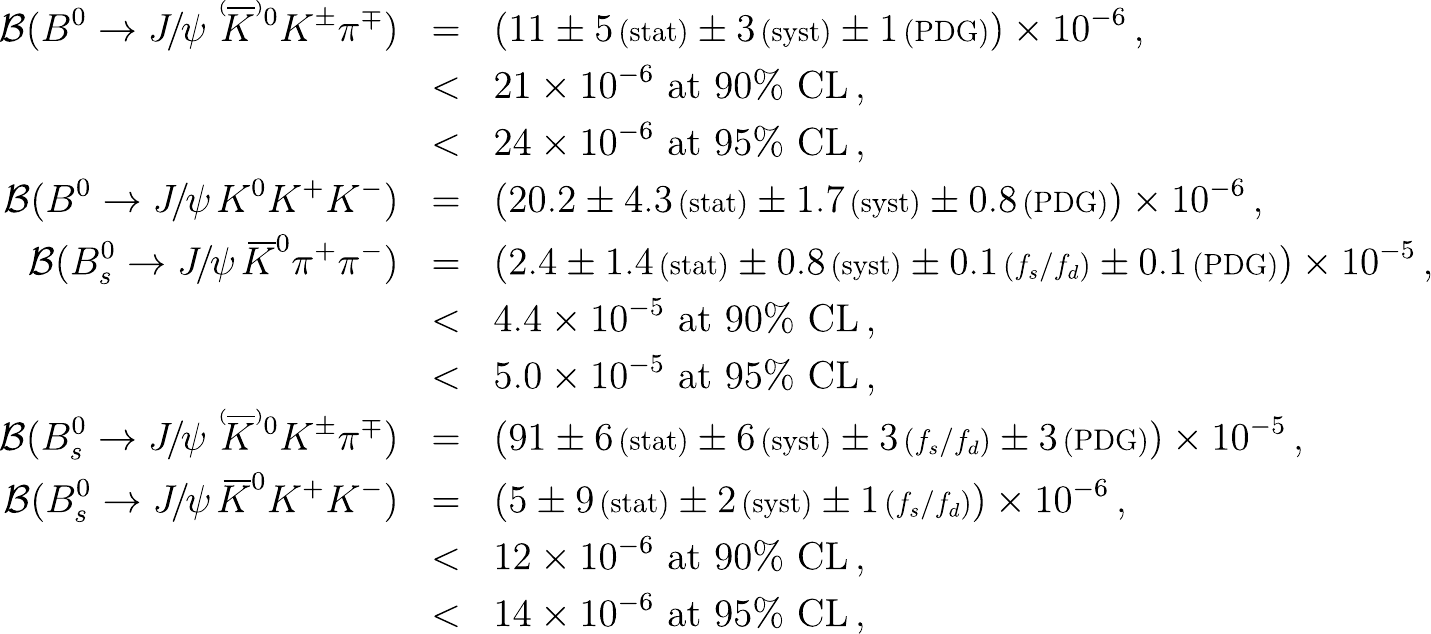}
\end{center}

\section{Epilogue}
While completing these proceedings, several important new results have been ob-
tained. LHCb updated the measurement of the CP-violating phase $\phi_s$ with 3\invfb~\cite{LHCb3fb}, and put limits on penguin effects in $\phi_s$~\cite{Aaij:2014vda}. The shift in $\phi_s$ is limited to be within the interval [-1.05$^\circ$, 1.18$^\circ$] at 95$\%$ C.L. HFAG collaboration updated the the $\phi_s$-$\Delta\Gamma_s$ plane~\cite{HFAGfall14}.

\Acknowledgements
I would like to thank the organizers of CKM2014 for the nice atmosphere during
the workshop in Vienna and my LHCb colleagues who helped in the preparation
of this talk.

\end{document}